\documentclass[]{spie}  

\usepackage{amsmath,amsfonts,amssymb}
\usepackage{graphicx}
\usepackage[colorlinks=true, allcolors=blue]{hyperref}

\title{The Acquisition Camera System for SOXS at NTT}

\author[a,c]{A. Brucalassi}
\author[b,c]{J. A. Araiza-Duran}
\author[c,b]{G. Pignata}
\author[d]{S. Campana}
\author[e]{R. Claudi}
\author[f]{P. Schipani}
\author[d]{M. Aliverti}
\author[e]{A. Baruffolo}
\author[h]{S. Ben-Ami}
\author[e]{F. Biondi}
\author[f]{G. Capasso}
\author[i,n]{R. Cosentino}
\author[j]{F. D'Alessio}
\author[d]{P. D'Avanzo}
\author[k]{D. Gardiol}
\author[g]{O. Hershko}
\author[l,m]{H. Kuncarayakti}
\author[n]{M. Munari}
\author[e]{D. Ricci}
\author[d]{M. Riva}
\author[g]{A. Rubin}
\author[n]{R. Zanmar Sanchez}
\author[n]{S. Scuderi}
\author[j]{F. Vitali}
\author[o]{J. Achr\'en }
\author[p]{I. Arcavi}
\author[d]{A. Bianco}
\author[e]{E. Cappellaro}
\author[f]{M. Colapietro}
\author[f]{M. Della Valle}
\author[g]{O. Diner}
\author[f]{S. D'Orsi}
\author[e]{D. Fantinel}
\author[q]{J. Fynbo}
\author[g]{A. Gal-Yam}
\author[d]{M. Genoni}
\author[r]{M. Hirvonen}
\author[l,m]{J. Kotilainen}
\author[m]{T. Kumar}
\author[d]{M. Landoni}
\author[r]{J. Lehti}
\author[s]{G. Li Causi}
\author[k]{D. Loreggia}
\author[e]{L. Marafatto}
\author[m]{S. Mattila}
\author[d]{G. Pariani}
\author[g]{M. Rappaport}
\author[e]{B. Salasnich}
\author[t]{S. Smartt}
\author[e]{M. Turatto}

\affil[a]{ESO-European Southern Observatory, Karl Schwarzschild Str. 2, D-85748, Garching bei Mu\"nchen, Germany}
\affil[b]{Millennium Institute of Astrophysics (MAS), Nuncio Monse\~nor S\'otero Sanz 100, Providencia, Santiago, Chile}
\affil[c]{Universidad Andres Bello, Av.da Republica 252, Santiago, Chile}
\affil[d]{INAF-Osservatorio Astronomico di Brera, Via Bianchi 46, I-23807, Merate (LC), Italy}
\affil[e]{INAF-Osservatorio Astronomico di Padova, Vicolo dell'Osservatorio 5, I-35122, Padua, Italy}
\affil[f]{INAF-Osservatorio Astronomico di Capodimonte, Salita Moiariello 16, I-80131, Naples, Italy}
\affil[g]{Weizmann Institute of Science, Herzl St 234, Rehovot, 7610001, Israel}
\affil[h]{Harvard-Smithsonian Center for Astrophysics, Cambridge, USA}
\affil[i]{INAF-Fundaci\'{o}n Galileo Galilei, Rambla J.A. Fernandez Perez 7, E-38712 Bre\~na Baja (TF), Spain}
\affil[j]{INAF-Osservatorio Astronomico di Roma, Via Frascati 33, I-00078 Monte Porzio Catone, Rome, Italy}
\affil[k]{INAF-Osservatorio Astrofisico di Torino, Via Osservatorio 20, I-10025 Pino Torinese (TO), Italy}
\affil[l]{FINCA-Finnish Centre for Astronomy with ESO, FI-20014 University of Turku, Finland}
\affil[m]{Department of Physics and Astronomy, University of Turku, FI-20014 Turku, Finland}
\affil[n]{INAF-Osservatorio Astrofisico di Catania, Via S. Sofia 78 30, I-95123 Catania, Italy}
\affil[o]{Incident Angle Oy, Capsiankatu 4 A 29, FI-20320 Turku, Finland}
\affil[p]{Tel Aviv University, Department of Astrophysics, 69978 Tel Aviv, Israel}
\affil[q]{DARK Cosmology Centre, Juliane Maries Vej 30, DK-2100 Copenhagen, Denmark}
\affil[r]{ASRO-Aboa Space Research Oy, Tierankatu 4B, FI-20520 Turku, Finland}
\affil[s]{INAF-Istituto di Astrofisica e Planetologia Spaziali, Via del Fosso del Cavaliere 100, 00133 Rome, Italy}
\affil[t]{Queen's University Belfast, County Antrim, BT7 1NN, Belfast, UK}

\authorinfo{Contact information: A.B.: E-mail: abrucala@eso.org, Telephone: +49 (0)89 3200 6456\\  
G.P.: E-mail: pignata@gmail.com, Telephone: +33 (0)1 98 76 54 32}

\begin{document} 
\maketitle

\begin{abstract}
SOXS (Son of X-Shooter) will be the new medium resolution (R$\sim$4500 for a 1 arcsec slit), high-efficiency, wide band spectrograph for the ESO-NTT telescope on La Silla. It will be able to cover simultaneously optical and NIR bands (350-2000nm) using two different arms and a pre-slit Common Path feeding system.
SOXS will provide an unique facility to follow up any kind of transient event with the best possible response time in addition to high efficiency and availability.
Furthermore, a Calibration Unit and an Acquisition Camera System with all the necessary relay optics will be connected to the Common Path sub-system. The Acquisition Camera, working in optical regime, will be primarily focused on target acquisition and secondary guiding, but will also provide an imaging mode for scientific photometry.\\ 
In this work we give an overview of the Acquisition Camera System for SOXS with all the different functionalities.
The optical and mechanical design of the system are also presented together with the preliminary performances in terms of optical quality, throughput, magnitude limits and photometric properties. 
 \end{abstract}


\section{INTRODUCTION}
\label{sec:intro}  
The \emph{ Son Of X-Shooter} (SOXS), is the new instrument\cite{2016Schipani} for the European Southern Observatory (ESO) New Technologies Telescope (NTT) at the La Silla Observatory, Chile, expected to have the first light at the end of 2020.
Largely based on the concept of X-Shooter\cite{2011Vernet} at the VLT,  SOXS will simultaneously cover the electromagnetic spectrum from 0.35 to 2.0$\mu$m with a spectral resolution of R$\sim$4500, using two arms\cite{soxsrubin, soxssanchez} (UV-VIS and NIR) and a pre-slit Common Path\cite{soxsclaudi} (CP) feeding system. 
A limiting magnitude of R=20mag point source with S/N$\sim$10 in a 1-hour exposure is foreseen.

The instrument will be dedicated to the characterization and classifications of all kind of transients sources, from predictable events like eclipses, transits, asteroids and comets, activity in young stellar objects, towards events with long reaction times, like supernovae, X-ray binaries
and novae, blazars magnetars, and AGN, up to events that need really fast reaction times, within one night or
less (electromagnetic counterparts of gravitational wave events, neutrino events, tidal disruptions of stars in the
gravitational field of supermassive black holes, gamma-ray bursts and fast radio bursts).

SOXS is also equipped with a Calibration sub-unit and an Acquisition Camera system (CAM) attached directly to the CP. The CAM system is foreseen to work not only for target acquisition and secondary guiding, but it will be also equipped with a filter wheel and a scientific camera to be used for some science grade imaging and moderate high speed photometry.

\subsection{The Acquisition Camera system}
\label{sec:CAM_intro}
   \begin{figure} [ht]
   \begin{center}
   \begin{tabular}{c} 
   \includegraphics[height=7cm]{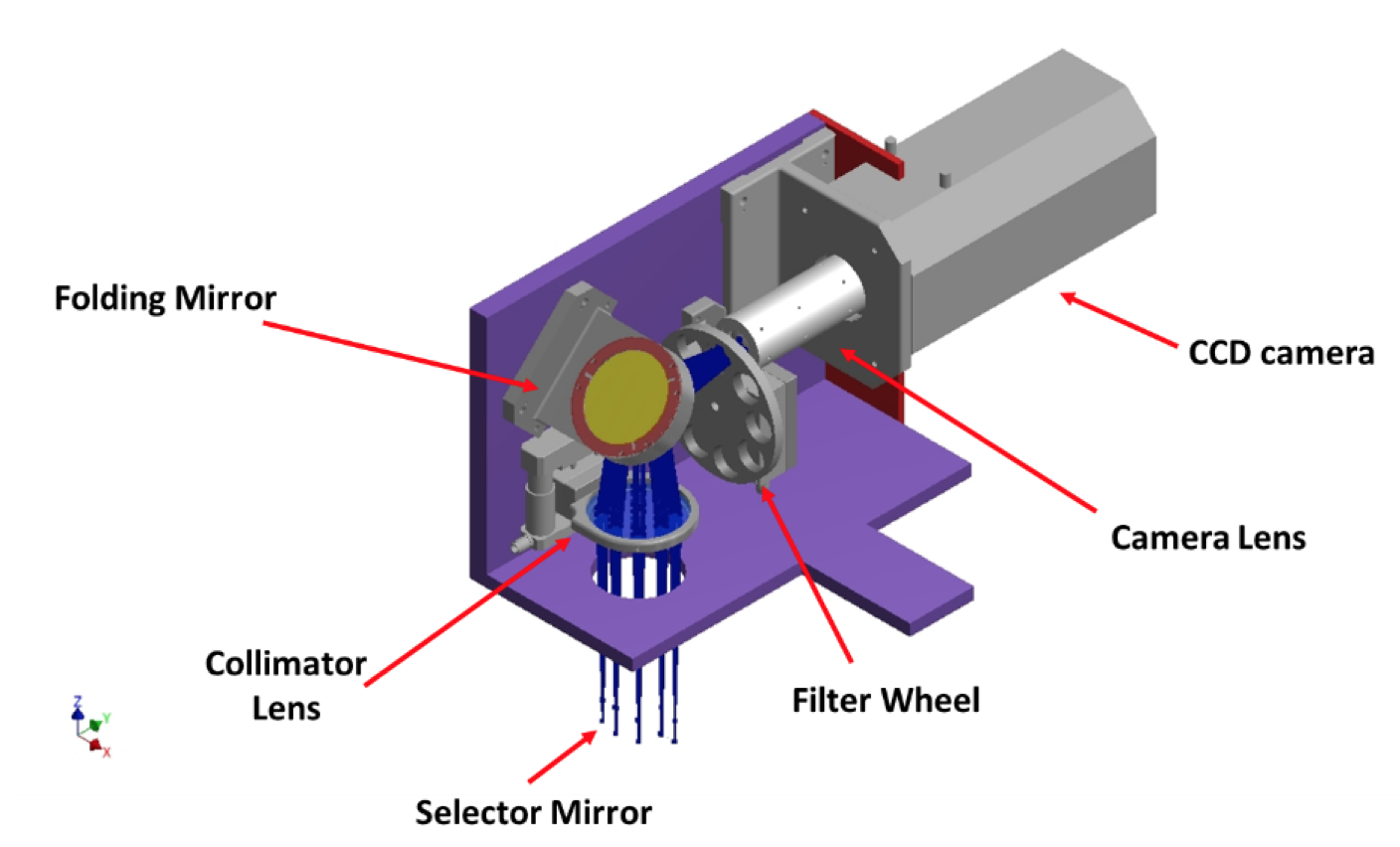}
   \end{tabular}
   \end{center}
   \caption[example] 
   { \label{fig:CAM_Overview}
Overview of the Acquisition Camera System:  it consists of a collimator lens, folding mirror, filter wheel, focal reducer optics and CCD camera.  
All the elements are included in a structure made of 6061-T6 Aluminum.}
   \end{figure}
\begin{figure} [ht]
   \begin{center}
   \begin{tabular}{c} 
   \includegraphics[height=7cm]{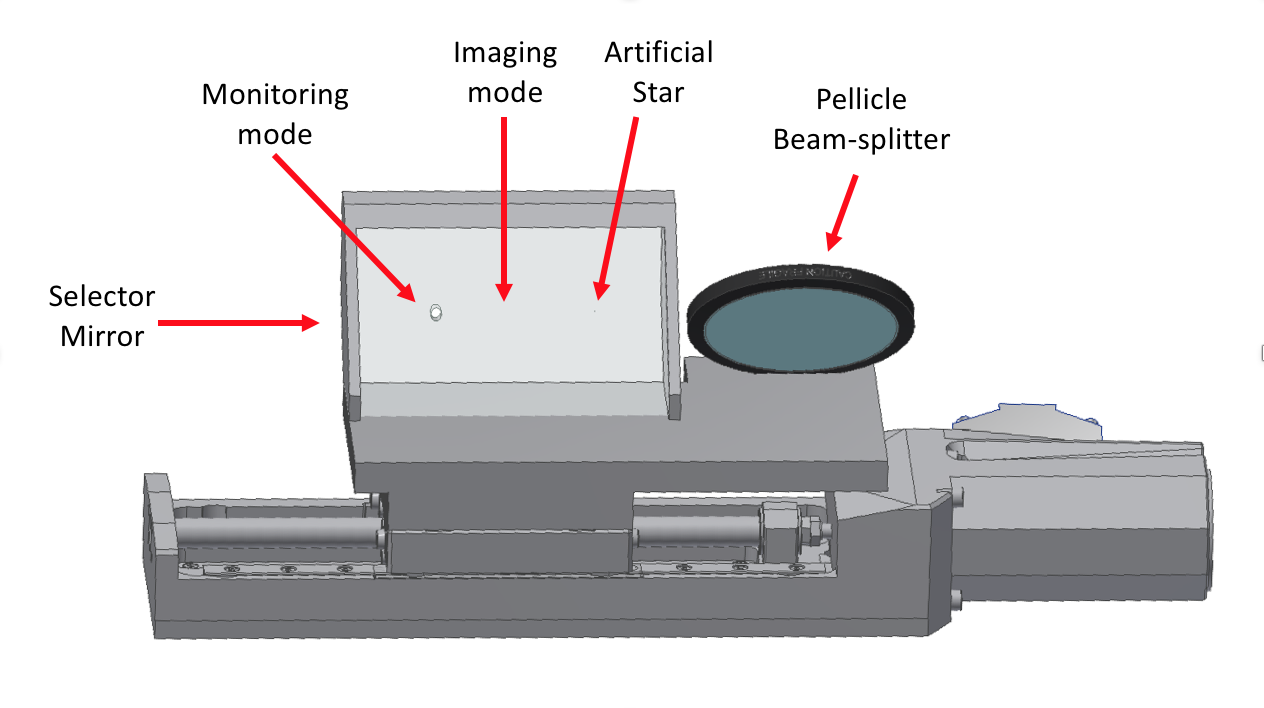}
   \end{tabular}
   \end{center}
   \caption[CAM_Selector] 
   { \label{fig:CAM_Selector} 
CAM selector system. A PI DC motor stage with 102mm stroke (PI L-406.40DD10)}
   \end{figure}

 
The Acquisition Camera system consists of a collimator lens, folding mirror, filter wheel, focal reducer optics and CCD camera.  
All the system is included in a structure made of 6061-T6 Aluminum. 
The CAM will receive an F/11 beam redirected from the telescope focal plane through the so called CAM selector placed in the CP\cite{soxsaliverti}.
The CAM selector (see Figure~\ref{fig:CAM_Selector}) is based on a sled that at the level of the Nasmyth focal plane, 
carries a single mirror with three positions for different functions and a pellicle beam-splitter. 
The mirror and the pellicle beam-splitter are tilted at 45$^\circ$ and direct light from sky or from the slits, respectively, to the CAM optics.
The focal plane is placed at 500mm from the NTT Nasmyth interface with a plate scale of 5.359 arcsec/mm. 
   
The main functionalities of the CAM system are described below, along with the corresponding CAM selector configurations:
\begin{description}
 \item[A:] \textbf{CP pellicle beam-splitter:}\\
 A semitransparent pellicle beamsplitter inclined towards the instrument allows to use the CAM system as a slit viewing camera (with calibration lamp on). 
Selecting different filters, one is able to use the CAM system as a selective on-slit viewer.
During daytime visually co-alignment of the slits could be possible by illuminating them with the calibration lamp through the pellicle.  
 
  \item[B:]\textbf{CP flat 45$^\circ$ mirror} with 3 selectable positions:
  \begin{enumerate}
   \item \textbf{ Acquisition and Imaging:}\\
   In this mode the 45$^\circ$ selector mirror on the CP will reflect the full field towards the CAM. 
   A complete broad-band filter set is available to aid during object acquisition and identification: 
   5 SDSS filters (u, g, r, i, z), Y to reproduce the LSST filter set, V (Johnson) and a free position for the slit viewer.
   This sub-system can also be employed as light imager in a given photometric band in order to perform photometry and flux calibration.   
    \item \textbf{Artificial Star:}\\
   The position is only used during day time maintenance.
   A 0.5 arcsec reference pinhole acts as an artificial star by switching on the calibration lamp. 
   Its proper centering in the spectrograph slits is verified with the science detectors. 
   The pinhole is also visible on the CAM and its position is recorded by the software as the reference for object centering.
     \item \textbf{Monitoring (Spectroscopy):}\\ 
     This mode will be selected during the science exposure. 
      The 45$^\circ$ mirror is translated to place a hole on the optical axis. 
      This passes an un-vignetted field of 15 arcsec (corresponding to 2.805mm) to the spectrograph slits.
      On the CAM it is possible to recover the peripheral field, but with a hole in the center.
      Off-axis secondary guiding on an object in the periphery will also be implemented during science exposures\cite{soxsricci} .
     \end{enumerate}
     
  \end{description}
This camera stage shall be able to place the mirrors with a maximum tilt of $\pm20$ arcsec and to position the holes with a repeatability of 0.5 micron.
In order to protect the delicate pellicle beamsplitter a protection will be provided.

\subsection{Detector System}
\label{sec:DetectorSystem}
A survey among manufacturers indicates the Andor iKon-M 934 Series Camera 
as a suitable detector for the CAM system. The selected CCD sensor model BEX2-DD assures a broadband coverage and a high NIR QE. 
In the following  the main characteristic of the camera are reported (For more details we refer to the Andor iKon-M 934 webpage
\footnote[1]{http://www.andor.com/scientific-cameras/ikon-xl-and-ikon-large-ccd-series/ikon-m-934}):
 \begin{description}
\item[$\bullet $] Active Pixels: 1024x1024, Pixel sixe: 13.0$\mu$m, Sensor Size: 13.3.x13.3 mm
\item[$\bullet $] Sensor Option: BEX2-DD: 
Back Illuminated, Deep Depletion with fringe suppression, extended range dual AR coating 
\item[$\bullet $] Pixel well depth: 130000e
\item[$\bullet $] Pixel readout rates: 5, 3, 1, 0.05 MHz, Frame Rate: 4.4 fps (full frame)
\item[$\bullet $] Read-out noise: 2.9e,  Dark current: 0.00012 e/pixel/sec at -100.0$^\circ$C
\item[$\bullet $] Maximum cooling: -100.0$^\circ$C \end{description}

\section{The CAM Optical Design}
\label{sec:CAM_Optical_Design} 
   \begin{figure} [ht]
   \begin{center}
   \begin{tabular}{c} 
   \includegraphics[height=8cm]{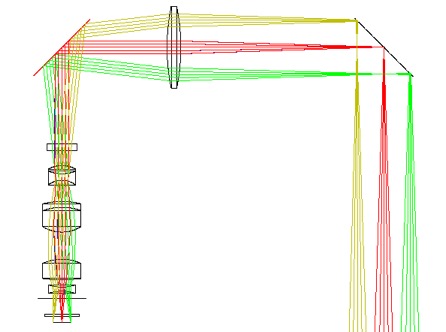}
   \end{tabular}
   \end{center}
   \caption[example] 
   { \label{fig:CAM_OpticalDesign} 
Optical Layout of the CAM System. From right to the left: the camera selector, the collimator lens (L1), the folding mirror, the filter wheel, the two doublets (D1, D2), the two singlets (S1, S2), the detector window and the focal plane are shown.}
   \end{figure} 
The following general requirements concerning the design and the performance of the CAM system have been specified:
\begin{description}
\item[$\bullet $]The CAM system shall ensure a Field of View of at least 2 arcmin x 2 arcmin (up to 4 arcmin x 4 arcmin).
\item[$\bullet $]Detector wavelength range shall extend up to 1.0-1.1 micron  with acceptable QE ($>20\%$).
\item[$\bullet $]Visually detect and obtain centroid position of object in u, g, r, i, z (Sloan) band, Y band to reproduce the LSST filter set, and V (Johnson) band.
\item[$\bullet $]Have a good image quality (with a scale smaller than 1 arcsec /1 pixel).
\item[$\bullet $]Use a 13.0$\mu$m , 1024x1024 pixel detector CCD optimized for NIR QE
\end{description}
Given these requirements, the obtained design is illustrated in Figure~\ref{fig:CAM_OpticalDesign}  and in Table~\ref{tab:OpticalElements_data} .
The largest Field of View that does not compromise the image quality was found to be 3.5x3.5 arcmin (linear), resulting in a pixel scale of 0.205 arcsec.\\ 
After the CAM selector, a lens of 60 mm diameter reimages the pupil onto a compact camera. In front of the camera there is a filter wheel, illustrated by a single plane glass. The subsequent camera, formed by 2 doublets and two singlets of max diameter 28 mm, relays the Nasmyth focus on the detector, with an $F_{\#}$=3.6.
The total length from focus to focus is 434.15 mm. The glasses are selected to provide colour correction over a wide wavelength range.
\begin{table}[ht]
\caption{Optical Elements data.} 
\label{tab:OpticalElements_data}
\begin{center}       
\begin{tabular}{|l|l|l|l|l|} 
\hline
\rule[-1ex]{0pt}{3.5ex}  Description & Radius(mm) & Distance (mm) & Material & Diameter(mm)\\
\hline
\rule[-1ex]{0pt}{3.5ex}  Telescope Focus & - & 147.66 & & 60.00   \\
\hline
\rule[-1ex]{0pt}{3.5ex}  L1 & 150.902 & 10.00 & CAF2 & 60.00  \\
\hline
\rule[-1ex]{0pt}{3.5ex}       & -147.935 & 76.50 &          & 60.00 \\
\hline
\rule[-1ex]{0pt}{3.5ex}  Folding Mirror & - & -70.00 &       & 46.00 \\
\hline
\rule[-1ex]{0pt}{3.5ex}    Filter              & - & -5.00 & N-BK7  & 22.00 \\
\hline
\rule[-1ex]{0pt}{3.5ex}                          &   & -10.97 &       &   \\
\hline
\rule[-1ex]{0pt}{3.5ex}    D1                      & -20.115  & -6.00 &  CAF2      & 20.00 \\
\hline
\rule[-1ex]{0pt}{3.5ex}                          & 26.793  & -4.82 & BAL15Y       & 20.00 \\
\hline
\rule[-1ex]{0pt}{3.5ex}                          & -16.688  & -15.42 &       & 20.00 \\
\hline
\rule[-1ex]{0pt}{3.5ex}    D2                     & -61.274  & -4.22 & PBL25Y      & 28.00 \\
\hline
\rule[-1ex]{0pt}{3.5ex}                          & -42.574  & -18.00 & CAF2      & 28.00 \\
\hline
\rule[-1ex]{0pt}{3.5ex}                          & 26.406  & -19.74 &       & 28.00 \\
\hline
\rule[-1ex]{0pt}{3.5ex}    S1                    & -32.969  & -15.01 &  BAL35Y     & 28.00 \\
\hline
\rule[-1ex]{0pt}{3.5ex}                          & 105.709  & -5.31 &       & 28.00 \\
\hline
\rule[-1ex]{0pt}{3.5ex}    S2                   & 29.584  & -3.00 & BSL7Y       & 20.00 \\
\hline
\rule[-1ex]{0pt}{3.5ex}                          & -31.441  & -5.00 &       & 20.00 \\
\hline
\rule[-1ex]{0pt}{3.5ex}                                  &   & -11.85 &       &  \\
\hline
\rule[-1ex]{0pt}{3.5ex}   Detector Window   &  - & -1.50 & SILICA       & 25.40 \\
\hline
\rule[-1ex]{0pt}{3.5ex}                                  &  & -4.15 &       &  \\
\hline
\rule[-1ex]{0pt}{3.5ex}  Focal Plane             & - &  &       & 14.00 \\
\hline

\end{tabular}
\end{center}
\end{table}

\subsection{Image Quality}
\label{sec:CAM_ImageQuality}
\begin{figure} [ht]
   \begin{center}
   \begin{tabular}{c} 
   \includegraphics[height=6cm]{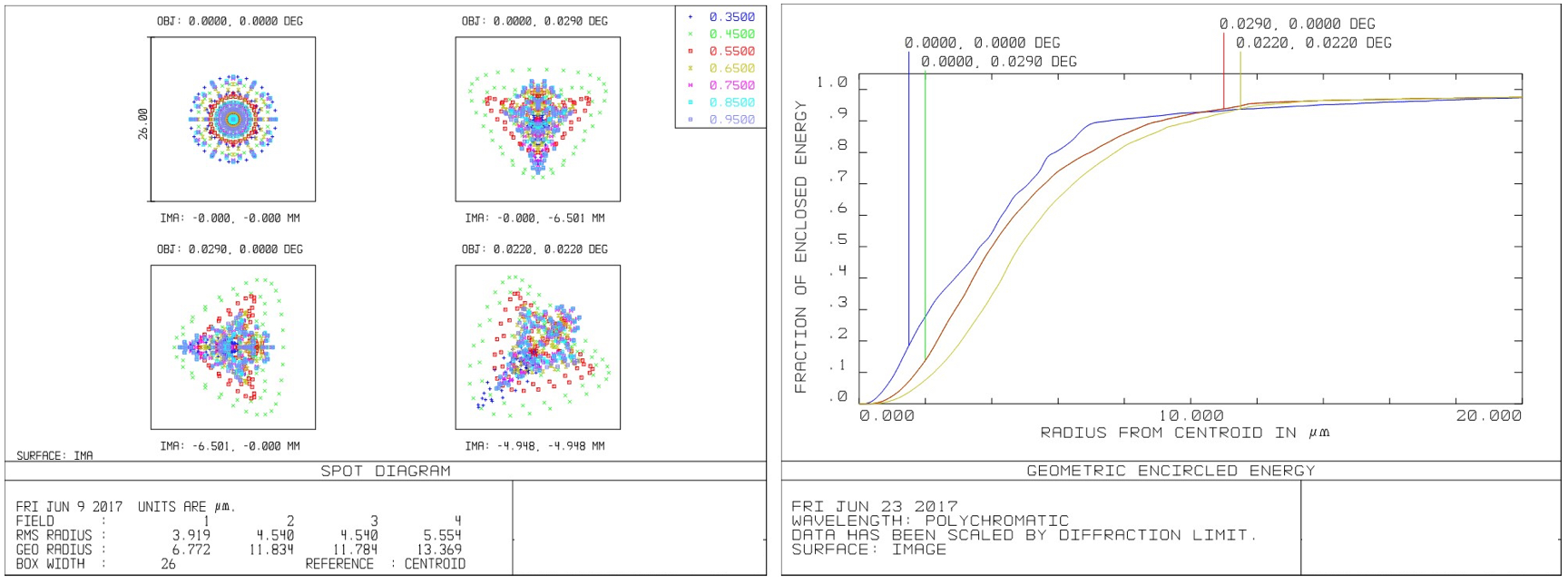}
   \end{tabular}
   \end{center}
   \caption[CAM_SpotDiagram] 
   { \label{fig:CAM_SpotDiagram} 
Left panel: CAM Spot diagrams for field positions correspond to on-axis, 3.5 arcmin (side of the detector) in x and y direction and 2.80x2.80 arcmin (corner of the detector).
Right panel: Geometric encircled  energy.}
   \end{figure}

Spot diagrams are reported in Figure~\ref{fig:CAM_SpotDiagram}, showing that the spot is contained in a box of 2x2 pixels (26.0 micron). 
Plotted field positions correspond to on-axis, 3.5 arcmin (side of the detector) in x and y direction and 2.80x2.80 arcmin (corner of the detector).
\subsection{Thermal Analysis}
\label{sec:CAM_Thermal Analysis}
\begin{figure} [ht]
   \begin{center}
   \begin{tabular}{c} 
   \includegraphics[height=6.2cm]{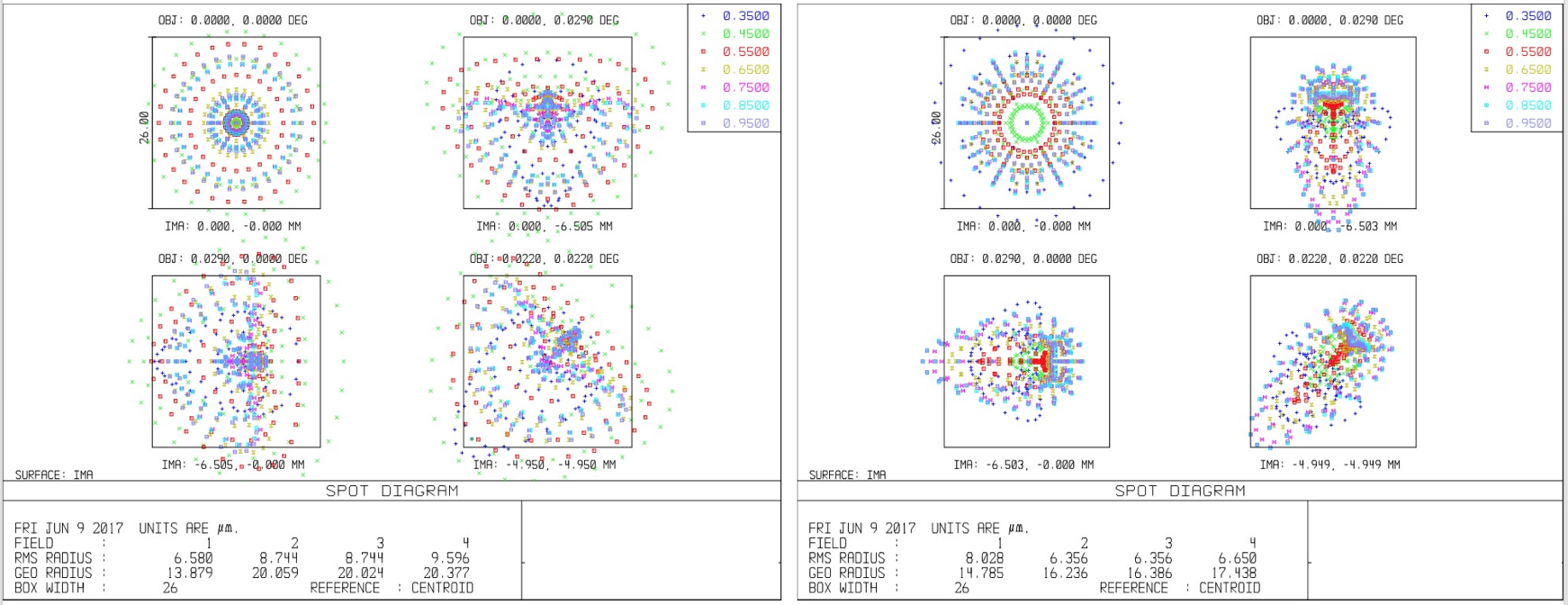}
   \end{tabular}
   \end{center}
   \caption[CAM_ThermalAnalysis] 
   { \label{fig:CAM_TempChanges_FDR} 
Spot diagrams considering temperatures of 0$^\circ$ C (left panel) and 20$^\circ$ C (right panel). The geometrical dimension of the spot are still about 2 pixels (26.0$\mu$m). }
   \end{figure}

The system has been designed at 10$^\circ$C. A change of $\pm10^\circ$C causes a deterioration of the image (see Figure~\ref{fig:CAM_TempChanges_FDR}), however, always with a geometrical dimension of about 2 pixels (26.0$\mu$m). In addition, the optical quality is completely recovered focusing the collimator lens (L1).
\subsection{Sensitivity Analysis}
\label{sec:CAM_Sensitivity Analysis}
The analysis of the expected performances for the CAM has been done using a set of of VIMOS and SDSS filters in the computations. The calculation has been made using an adapted version of the algorithm from "Astronomical Computing" Sky \& Telescope, May 1998 by Bradley E. Schaefe, with general parameters for the system summarized in Table~\ref{tab:GeneralParameters}.
\begin{table}[ht]
\caption{General Parameters for Sensitivity Prediction} 
\label{tab:GeneralParameters}
\begin{center}       
\begin{tabular}{|l|l|}
\hline
\rule[-1ex]{0pt}{3.5ex}  \textbf{Parameter} & \textbf{Value }  \\
\hline
\rule[-1ex]{0pt}{3.5ex}  Telescope Aperture & 350 cm   \\
\hline
\rule[-1ex]{0pt}{3.5ex}  Secondary Mirror Diameter  & 87.5 cm   \\
\hline
\rule[-1ex]{0pt}{3.5ex}  Number of Mirror surfaces & 5  \\
\hline
\rule[-1ex]{0pt}{3.5ex}  Number of lenses in light path & 5\\
\hline
\rule[-1ex]{0pt}{3.5ex}  Pixel scale & 0.20 arcsec/pixel  \\
\hline 
\rule[-1ex]{0pt}{3.5ex}  RON & 2.9e  \\
\hline 
\rule[-1ex]{0pt}{3.5ex} Number of exposure & 1   \\
\hline 
\rule[-1ex]{0pt}{3.5ex}  Airmass & 1.15   \\
\hline 
\rule[-1ex]{0pt}{3.5ex}  FWHM seeing disk & 0.8 arcsec   \\
\hline
\rule[-1ex]{0pt}{3.5ex}  Temperature & 10$^\circ$ C  \\
\hline 
\rule[-1ex]{0pt}{3.5ex}  Relative Humidity & 10\%  \\
\hline 
\rule[-1ex]{0pt}{3.5ex}  Observatory & La Silla (Chile)   \\
\hline 
\rule[-1ex]{0pt}{3.5ex}  Moon Phase & New   \\
\hline 
 
\end{tabular}
\end{center}
\end{table}

The filter peak transmission has been considered to be 80.0\% for all the photometric bands, while the quantum efficiency of the detector 
is taken from the Andor iKon-M934 datasheet (see Sec.~\ref{sec:DetectorSystem}).
The computation of the limiting magnitude as function of the exposure time has been done for SNR=10. The results for SDSS are reported in the Table~\ref{tab:Limiting Magnitude}.
\begin{table}[ht]
\caption{Limiting Magnitude of the CAM system for a SNR=10 in SDSS band.} 
\label{tab:Limiting Magnitude}
\begin{center}       
\begin{tabular}{|l|l|l|l|l|l|l|l|}
\hline
\rule[-1ex]{0pt}{3.5ex}  \textbf{SDSS Band} &   \textbf{1 sec} &  \textbf{2 sec} & \textbf{3 sec} &  \textbf{5 sec }& \textbf{10sec} &  \textbf{15 sec} &   \textbf{20 sec}   \\
\hline
\rule[-1ex]{0pt}{3.5ex}  u' (355.7nm) & 15.9 & 16.7 &  17.5	 & 17.7 & 18.4 & 18.7 & 19.1  \\
\hline
\rule[-1ex]{0pt}{3.5ex}  g' (482.5nm) &  18.2& 18.9 &	19.4 &19.8 &	20.5	&20.8&	21.0 \\
\hline
\rule[-1ex]{0pt}{3.5ex}  r' (626.1nm)& 18.0&	18.6&	19.0&	19.5&	20.0&	20.3&	20.4  \\
\hline
\rule[-1ex]{0pt}{3.5ex}  i' (726.2nm)&  16.4&	17.1&	17.5&	17.9&	18.4&	18.6&	18.8 \\
\hline
\rule[-1ex]{0pt}{3.5ex}  z' (909.7nm)& 15.3&	15.9&	16.2&	16.5&	16.9&	17.2&	17.4 \\
\hline 
\end{tabular}
\end{center}
\end{table}

\section{The CAM Mechanical Design}
\label{sec:CAM_Mechanical_Design} 

The CAM has a main T-shape structure made by Aluminum 6061-T6 as shown in Figure~\ref{fig:CAM_Overview}. To ensure the maximum stiffness all the walls are structural with thickness of around 15 mm. The weight of this part is about 1.4 kg. The cover is not structural and is made of a 3 mm thick Aluminum plate.
An aluminum support will be mounted on the structure to fix the detector. This operation will be performed screwing it directly on the front surface of the structure.  The centering of the CCD will be done shimming the 3 holes and 3 dowel pins used to place the CCD together with the support system.

\subsection{Opto-mechanical elements}
\label{sec:CAM_Elements}
\begin{figure} [ht]
   \begin{center}
   \begin{tabular}{c} 
   \includegraphics[height=6cm]{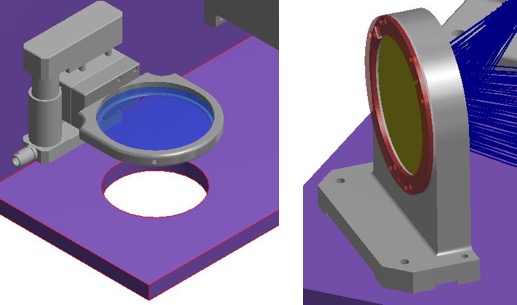}
   \end{tabular}
   \end{center}
   \caption[CAM_OptomechElem] 
   { \label{fig:CAM_OptomechElem} 
Left panel: Preliminary mechanical design of the Collimator Lens support. Also the linear stage is visible. Right panel: Preliminary mechanical design of the Folding Mirror support.  }
   \end{figure}

Following the optical path, after the CAM selector, a 60 mm diameter lens (collimator) is foreseen to act as a re-focuser and will be placed on a linear stage with 15 mm stroke (PI M-111.1DG1). The lens is glued to an Aluminum mount by 3 3M 2216 glue spots (see Figure~\ref{fig:CAM_OptomechElem}, left panel). 
The beam is redirected through a folding mirror (see Figure~\ref{fig:CAM_OptomechElem}, right panel) along an axis parallel to the main telescope axis. 
The mirror will be mounted on a support screwed on the front surface of the external structure and kept in position inside the mount frame by axial leaf springs and a spring retainer. In front of the camera lens there is a filter wheel made by a rotary stage (PI M-116.DG) and a custom filter support which can host up to 9 elements.
   \begin{figure} [ht]
   \begin{center}
   \begin{tabular}{c} 
   \includegraphics[height=7cm]{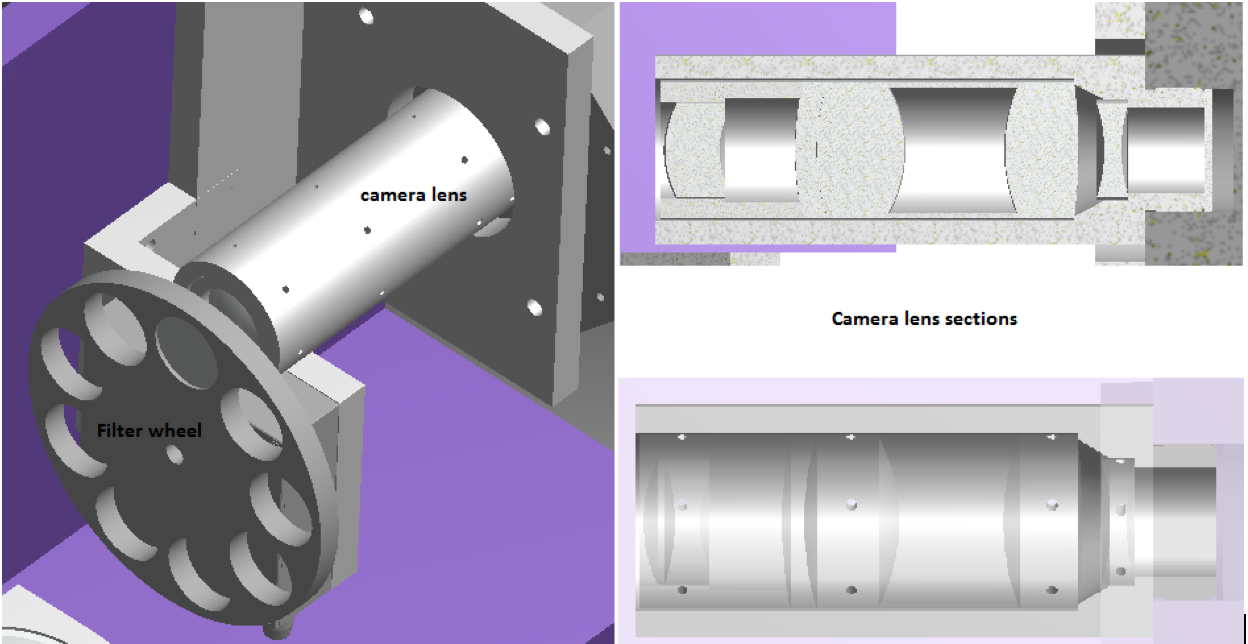}
   \end{tabular}
   \end{center}
   \caption[CAM_CamLens] 
   { \label{fig:CAM_CamLens} 
Left: CAM camera lens and filter wheel in isometric view. Right: section of the camera tube: the 2 doublets and the 2 singlets are also visible.}
   \end{figure} 
The subsequent camera, formed by 2 doublets and two singlets of max diameter of 28 mm, relays the Nasmyth focus on the detector, with a F$_{\#}$=3.6 (see Figure~\ref{fig:CAM_CamLens}). The total length form focus to focus is 434.15 mm. For the camera lens, it is foreseen to include the 2 doublets and the 2 singlets in a tubular structure anchored to the camera mounting. The optical components will be mounted creating 3 3M 2216 glued spots on the side of the element. This operation will be performed via small holes radially drilled inside the support.

 \section{CONCLUSION}
\label{sec:CAM_conclusion}
We presented an overview of the Acquisition Camera system for the NTT instrument SOXS.
The compact optical design and a scientific CCD camera optimized for NIR QE ensure a FOV of 3.5x3.5 arcmin and good quality image (with a pixel scale of 0.205 arcsec/pixel), such that the CAM system is foreseen to work not only for target acquisition and secondary guiding, but it will be also used for some science grade imaging and moderate high speed photometry.
The CAM system will be included in T-shape structure made by Aluminum 6061-T6 with a no-structural cover made of a 3 mm thick Aluminum plate. The total sub-system will be anchored to the CP through kinematic mounts.


\bibliography{report} 
\bibliographystyle{spiebib} 

\end{document}